\documentclass[aip, 12pt,showkeys]{revtex4-2} 	
\usepackage{natbib}
\usepackage[unicode]{hyperref}
\usepackage{dcolumn}
\usepackage{bm}
\usepackage{graphicx}
\usepackage{upgreek}
\usepackage[dvipsnames]{xcolor}
\usepackage[normalem]{ulem}
\usepackage{amsmath}
\usepackage[utf8x]{inputenc}
\usepackage{braket}

\newcommand{\newartem}[1]{\textcolor[HTML]{990AB6}{#1}}

\newcommand{\clocktransitionfirst}{$\ket{g,F=4,m_F=0}\rightarrow \ket{c,F=3,m_F=0}$\space}
\newcommand{\clocktransitionsecond}{$\ket{g, F=3,m_F=0}\rightarrow \ket{c,F=2,m_F=0}$\space}

\newcommand{\clockw}{$1.14\,\mu$m\space}
\newcommand{\bluefirst}{$\ket{g,F=4}\rightarrow\ket{b,F=5}$\space}
\newcommand{\bluesecond}{$\ket{g,F=3}\rightarrow\ket{b,F=4}$\space}
\newcommand{\clockfirstshort}{''4-3''\space}
\newcommand{\clocksecondshort}{''3-2''\space}

\begin{document}

\title{Extraordinary low systematic frequency shifts in bi-colour thulium optical clock}

\author{A.\,Golovizin,$^{1}$}\email{artem.golovizin@gmail.com}, 
\author{D.\,Tregubov,$^{1}$
E.\,Fedorova,$^{1}$
D.\,Mishin,$^{1}$
D.Provorchenko,$^{1}$
K.\,Khabarova,$^{1,\,2}$ V.\,Sorokin,$^{1}$ N.\,Kolachevsky$^{1,\,2}$}

\affiliation{
$^1$\,P.N.\,Lebedev Physical Institute, Leninsky prospekt 53, 119991 Moscow, Russia\\
$^2$\,Russian Quantum Center, Bolshoy Bulvar 30,\,bld.\,1, Skolkovo IC, 121205 Moscow, Russia\\
}
\date{\today}

\begin{abstract}
Optical atomic clocks have already overcome the eighteenth decimal digit  of  instability and uncertainty  demonstrating incredible control over external perturbations of the clock transition frequency \cite{Huntemann2016,McGrew2018atomic,brewer1029systematic,bothwell2019JILA,Schioppo2016,Oelker2019demonstration}.
At the same time there is an increasing demand for atomic and ionic transitions with minimal sensitivity to external fields, with practical operational wavelengths and robust readout protocols \cite{campbell2012single,Ohtsubo2019optical,Arnold2018,micke2020coherent}. 
One of the goals is to simplify clock's operation maintaining  its relative uncertainty at low $10^{-18}$ level.  It  is especially important for transportable and  envisioned space-based optical clocks \cite{Delva2017a,Schiller2017space}.
We proved earlier \cite{Sukachev2016,Golovizin2019inner} that the  \clockw inner-shell magnetic dipole transition in neutral thulium possesses very low blackbody radiation shift compared to other neutrals.
Here we demonstrate operation of a bi-colour thulium optical clock with extraordinary low sensitivity to the Zeeman shift due to a simultaneous interrogation of two clock transitions  and data processing. Our experiment shows suppression of the quadratic Zeeman shift by at least three orders of magnitude.
The effect of tensor lattice Stark shift can be also reduced to below $10^{-18}$ in fractional frequency units.
All these features make thulium optical clock almost free from hard-to-control systematic shifts. 
Together with convenient cooling and trapping laser wavelengths, it  provides  great perspectives for thulium lattice clock as a high-performance transportable system.
\end{abstract}

\maketitle
\label{Section:intro}
Excellent applicability of atoms as frequency references  owes to high-quality factor of certain atomic transitions and possibility to isolate atoms from the environment.
The first atomic microwave frequency reference \cite{essen1955atomic} was demonstrated in 1955, and in 1967 the SI second was redefined \cite{terrien1968news}  (9.2\,GHz hyperfine transition in Cs). 
Later, advances in laser technology,  manipulation of atoms and ions and optical frequency measurements  paved a way to high-Q optical frequency references \cite{Poli2014, Ludlow2015, udem2002optical}. 
 Modern  optical clocks operating at  $10^{-18}$ level of relative instability and uncertainty \cite{Ohmae2020direct,bacon2020frequencyarxiv,Oelker2019demonstration,milner2019demonstration}
stimulate discussion about inevitable redefinition of the SI second  \cite{Riehle2018,lodewyck2019definition}, which is also  motivated by comparisons of different types of optical clocks at   $10^{-17}$ level \cite{Nemitz2016,Ohmae2020direct, bacon2020frequencyarxiv, Dorscher2021optical}. Such  performance allows to accurately test some fundamental physical theories: general relativity \cite{Takamoto2020test,Grotti2018}, Lorentz invariance \cite{Sanner2019optical,Lange2021improved}, drifts of fundamental constants \cite{Huntemann2014,godun2014frequency}, search for the dark matter particles \cite{Kennedy2020precision} and dark matter clusters \cite{Roberts2020search}.

High-performance transportable optical clocks are considered as an essential part of a worldwide optical clocks network \cite{Grotti2018, Riehle2017}. 
Together with stabilized optical fibre links \cite{bacon2020frequencyarxiv,Lisdat2016network}, transportable systems are requested for long-distant time and frequency comparisons and  chronometric levelling \cite{Grotti2018,Takamoto2020test}. The most advanced systems are based on single ions in Paul traps and ensembles of neutral atoms in optical lattices.
Single-ion systems are typically less sensitive to the environment and allow simpler setups, while  clocks based on ensembles of neutral atoms (typically up to $10^5$) demonstrate a better stability. 
Largest systematic shifts in lattice clocks  come from  the lattice and interrogation light fields,  blackbody radiation (BBR), and an external  magnetic field (the Zeeman shift) \cite{poli2014transportable, Grotti2018,Takamoto2020test}. 
In this paper we demonstrate that all above mentioned shifts can be reduced to $10^{-18}$ level or lower in thulium bi-colour lattice optical clock.

\begin{figure}[h!]
\center{
\resizebox{0.5\textwidth}{!}{
\includegraphics{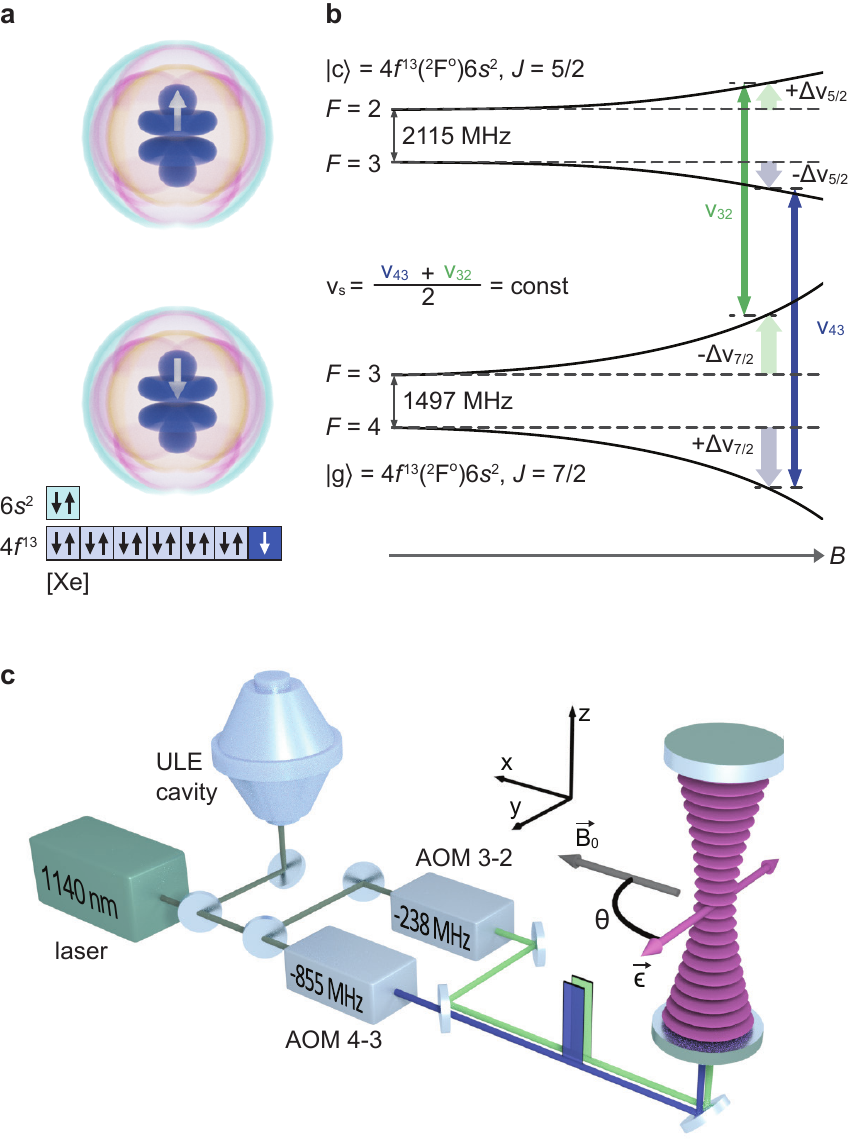}
}
}
\caption{
Bi-color  interrogation scheme in Tm clock using the hyperfine components of the clock levels.
a) Visual representation of electronic structure of the ground and clock levels.
Unpaired 4f-electron is located inside of the closed 6s shell. 
White arrow represents spin orientation of the electron.
b) magnetic doublets of the ground- and excited-states clock levels. Two simultaneously interrogated transition frequencies are denoted as $\nu_{43}$ and $\nu_{32}$. 
c) sketch of the experimental setup. The \clockw laser is stabilized to ULE cavity. Two acousto-optical modulators (AOMs) form pulse sequences  for the simultaneous ''4-3'' and ''3-2'' clock transition\newartem{s} interrogation of atoms trapped in the vertically-oriented optical lattice at 1064\,nm. 
}
\label{fig:scheme}
\end{figure}

The advantageous metrological properties of the \clockw inner shell clock transition in $^{169}$Tm are associated with similarity of the clock levels and their wave functions. 
The clock levels  are the fine-structure components of the same  ground state  electronic level separated by the  optical interval of 1.14 $\mu$m.
The transition is associated with a spin flip in the  $4f$-electron shell, which is strongly shielded  by the outer closed $5s^2$ and $6s^2$ shells (Fig.\,\ref{fig:scheme}a).
Both factors make the transition frequency highly insensitive to the {\it{dc}} external electric field and collisions, which was first experimentally observed already in 1983  \cite{aleksandrov19831}.
We showed previously \cite{Sukachev2016,Golovizin2019inner} that the \clockw ($\nu_0=2.6\times10^{14}$\,Hz) transition possesses very small susceptibility to a {\it{dc}} external electric field which provides  very low sensitivity to black body radiation (BBR). At the room temperature the BBR frequency shift  corresponds  to only $2.3(1.1)\times10^{-18}$ in fractional units which is thousands of times smaller than, e.g. in strontium lattice clock.
We also observed  close-to-zero differential polarizability of the clock levels of $0.00\,(0.01)$\,a.u.  at 1064\,nm, which indicates the proximity of the magic wavelength in a very practical spectral region where powerful  and low-noise fiber lasers are available.  
This magic wavelength has a number of useful features. The polarizability slope of
 $10^{-3}$\,a.u./nm   at 1064\,nm  is three orders of magnitude smaller than for $^{87}$Sr  at the magic wavelength 813\,nm \cite{Ushijima2018operational}.
This significantly softens the requirements to the lattice laser frequency stabilization (GHz instead of MHz range). Large detuning from any transition frequency in Tm and the absence of the  close-to-resonant terms in hypolarizability \cite{Sukachev2016} should also result in much smaller 
multipolar polarizability $\Delta\alpha_{M1+E2}$ and hyperpolarizability $\beta$ compared to  $^{87}$Sr. 
Note, that at the same time in Sr optical clocks the lattice shifts can be suppressed to the low $10^{-18}$ level \cite{Ushijima2018operational,Nicholson2015systematic}.

Yet the advantages of the $4f-$shell clock transition come with a price: asymmetric structure of the electron wave-function and the strong magnetic dipole-dipole interaction. 
The contribution from  the dipole-dipole interaction can be readily cancelled out by use of transition between zero projections of total atomic momentum  $m_F=0\rightarrow m_F=0$.
The asymmetric wave function results in a non-zero differential tensor polarizability which is numerically small ($0.2$\,a.u. at 1064 nm), but significant at the requested level of uncertainty. Still, the  key systematic contribution to the frequency shift of the $4f-$shell clock transition is the second order Zeeman shift of 257\,Hz/G$^2$ which is large compared to other atomic species used in optical clocks. 
Exactly these two effects --- the Zeeman effect and the tensor polarizability ---  in our case account for  the main undesired systematic contributions.
Handling these two shifts simultaneously requires unconventional  approach because of  their opposite behaviour with respect to the bias magnetic field $B_0$. Decreasing of the bias magnetic field reduces the quadratic Zeeman shift and corresponding uncertainty, while uncertainty of the tensor Stark shift generally increases.

Here we propose and experimentally  implement  a method for complete cancellation of the Zeeman shift and for the tight  control over the tensor polarizability shift at lower than 1\,mHz level which corresponds to $3.8\times10^{-18}$ in relative frequency units. 
This approach provides extremely low, compared to other widespread optical clocks, net systematic frequency shift and corresponding uncertainty in Tm system at $10^{-18}$ level. 
The idea is to use simultaneous  bi-colour interrogation scheme of two ground-state hyperfine sublevels $F=4$ and $F=3$ which allows to use the instantaneous synthetic frequency
\begin{equation}
    \nu_\textrm{s} = \frac{\nu_{43} + \nu_{32}}{2} 
\end{equation}
as depicted in Fig.\,\ref{fig:scheme}b. Here  $\nu_{43}$ and $\nu_{32}$ are the frequencies of the  \clocktransitionfirst (short notation \clockfirstshort) and \clocktransitionsecond (\clocksecondshort) transitions, respectively. 
Here $\ket{g}$ stands for the $\ket{4f^{13}(^2F^o)6s^2,J=7/2}$ ground level and $\ket{c}$ stands for the $\ket{4f^{13}(^2F^o)6s^2,J=5/2}$ clock level.
It turns out that  the synthetic frequency $\nu_\textrm{s}$ is  completely insensitive to the  magnetic field because of  equal but opposite Zeeman shifts of \clockfirstshort and \clocksecondshort clock transitions (see Supplementary for details). 

Synthetic frequency is  used on a regular basis in various  optical clocks when different magnetic sublevels are interrogated successively  \cite{bernard1998}.
However proposed approach possesses an important advantage: in our configuration we simultaneously probe two hyperfine clock transitions, such that the impact of fluctuating external magnetic field on two frequencies becomes completely identical. 
This fact  results in a full cancellation of the Zeeman shift without any assumptions about magnetic field behaviour between consecutive measurements.
To successfully implement this approach  one should  (i) simultaneously  prepare an atomic ensemble in two initial states, and (ii)  independently and simultaneously interrogate and  readout two hyperfine clock  transitions every measurement cycle. 

Simultaneous single-frequency optical pumping to $\ket{F=3,4\,, m_F=0}$ central magnetic sublevels was  demonstrated in our previous work  \cite{fedorova2020simultaneous}. In turn, relatively small separation of the clock transition frequencies of $\Delta\nu = \nu_{32} - \nu_{43} \approx 617$\,MHz simplifies  simultaneous interrogation and readout using  acousto-optical modulators (AOMs). 
At the same time, the {\it{ac}} Stark shift induced by the bi-colour probe field is only $0.06\,\mu$Hz in our experimental conditions because of  the low probe field intensity. 
The mutual influence of interrogation and readout fields in the bi-colour scheme is discussed in Supplementary.
It is also worth mentioning that polarizabilities of $m_F=0$ sublevels in the  hyperfine doublets are identically equal such that  the condition for the magic wavelength fulfils simultaneously for both clock transitions.

\textbf{Zeroing the Zeeman shift in the bi-colour scheme }

We prepare approximately $10^5$ atoms in the $\ket{g,F=4,m_F=0}$ state and $10^4$ atoms in the $\ket{g,F=3,m_F=0}$ state using simultaneous optical pumping \cite{fedorova2020simultaneous} with a following lattice depth ramp to sift out hot atoms from the trap.
Two clock transitions are simultaneously excited by $80$\,ms-long clock laser $\pi-$pulses with the frequencies $\nu_{43}$ and $\nu_{32}$  separated by approximately 617\,MHz. 
Using the readout procedure described in Supplementary,  we simultaneously  measure the excitation probability for each of the transition for certain probe field detunings. 
We  prove that  our experimental procedure allows to independently deduce two  excitation efficiencies. 
The system operates in the classical optical clock regime, when the laser frequency is alternatively switched between the left and the right slope of the clock transition. 
In our case the spectral linewidth of both clock transitions is $\delta\nu=10$\,Hz. 
Using this method we deduce two independent error signals and two clock transition frequencies  $\nu_{43}$ and $\nu_{32}$. 
As a stable frequency reference we use 1.14\,$\mu$m clock cavity with the linear drift correction \cite{Golovizin2019ultrastable}.
The impact of an external parameter (e.g. the bias magnetic field $B_0$) on the transition frequency can be studied by differential measurement  when the parameter is changed alternatively between two values for odd and even measurement cycles. The multi-channel digital lock tracks corresponding transition frequencies.
Calibrating measurements are periodically performed in order to monitor and tune auxiliary parameters (see Supplementary).
  
\begin{figure}[t]
\center{
\resizebox{0.45\textwidth}{!}{
\includegraphics{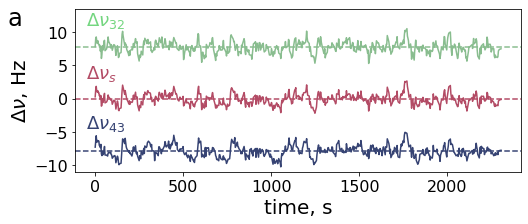}
}\\
\resizebox{0.45\textwidth}{!}{
\includegraphics{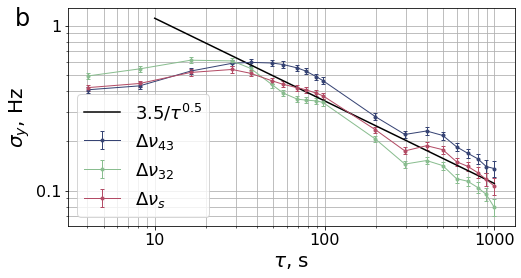}
}
\caption{a) Measurement of clock transitions frequency shifts $\Delta \nu_{43}=\nu_{43}(B^\textrm{m})-\nu_{43}(B^\textrm{r})$ (blue) and $\Delta \nu_{32}=\nu_{32}(B^\textrm{m})-\nu_{32}(B^\textrm{r})$ (green) for two alternatively changing values of the bias magnetic field $B^\textrm{r}_0=218$\,mG and  $B^\textrm{m}_0=132$\,mG. Common features in $\Delta \nu_{43}$ and $\Delta \nu_{32}$ data sets come from  clock laser frequency fluctuations. 
The synthetic frequency shift $\Delta\nu_\textrm{s}=(\Delta\nu_{43}+\Delta\nu_{32})/2$ is plotted in red with the mean value of $-0.08(6)$\,Hz (dashed line).
b) The Allan variance plotted for each of the data sets. The solid black line indicates  $\sim1/\sqrt{\tau}$ dependency.
\label{fig:shifts}}
}
\end{figure}

In order to  experimentally verify cancellation of the Zeeman shift for the synthetic frequency $\nu_\textrm{s}$, we  change the bias magnetic field between two largely different  values: the ``reference'' field $B^\textrm{r}_0=218$\,mG and the ``main'' field  $B^\textrm{m}_0=132$\,mG. The digital locks track corresponding frequency shifts with the help of the AOMs (see Fig.\ref{fig:scheme}c). We deduce differential frequency shifts of two clock transitions 
\begin{equation}\label{eq:dnus1}
    \begin{split}
        \Delta\nu_{43} &= \nu_{43} (B^\textrm{m}) - \nu_{43}(B^\textrm{r})\,, \\
    \Delta\nu_{32} &= \nu_{32}(B^\textrm{m}) - \nu_{32}(B^\textrm{r}) \\
    \end{split}
\end{equation}
and the corresponding synthetic frequency shift 
\begin{equation}\label{eq:dnus2}
    \Delta\nu_\textrm{s} = (\Delta\nu_{43} + \Delta\nu_{32})/2.
\end{equation}

The frequency differences  $\Delta\nu_{43}$, $\Delta\nu_{32}$, and $\Delta\nu_\textrm{s}$ measured for 2300 s are shown in Fig.\,\ref{fig:shifts}a.  
The mean value for $\Delta\nu_{43}$ equals to $-7.88(9)$\,Hz, while for $\Delta\nu_{32}$ it equals to $+7.72(7)$\,Hz, where the number in parentheses indicates one standard deviation. 
For the synthetic frequency the effect  averages out to $-0.08(6)$\,Hz. 
Figure \ref{fig:shifts}b shows the Allan variance for the  corresponding data sets. 
For longer averaging times we see the $3.5\times\tau^{-1/2}$\,Hz behaviour for $\Delta\nu_\textrm{s}$ which corresponds to $1.4\times 10^{-14}\tau^{-1/2}$ (here $\tau$ is taken in seconds) in relative units. For short averaging times of 1 s the Allan variance approaches instability of the clock laser of $2\times10^{-15}$ which was characterized earlier \cite{Zalivaco2020compact}. 
Fluctuations of interrogation laser  and the performance of digital locks are discussed in Supplementary.

\begin{figure}
\center{
\resizebox{0.5\textwidth}{!}{
\includegraphics{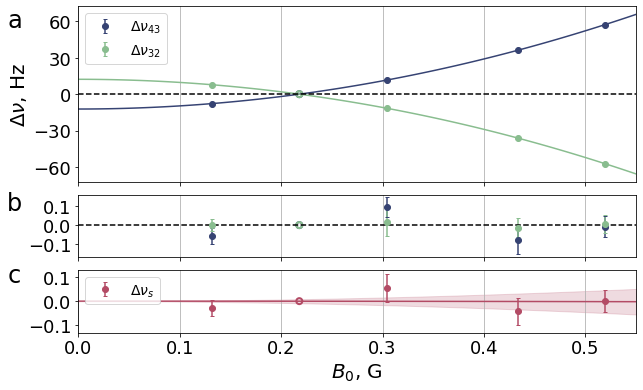}
}
\caption{Zeeman frequency shifs. a) Frequency shifts $\Delta\nu_{43}$ and $\Delta\nu_{23}$ as a function of the ``main'' bias magnetic fields $B^\textrm{m}_0$ at the fixed value of the ``reference'' field of $B^\textrm{r}_0=218$\,mG. 
Solid lines are the  parabolic fits to the data. 
b) The fit residuals. 
The synthetic  frequency shift $\Delta\nu_\textrm{s}$ is shown in~c). The  shaded area  indicates $1\sigma=0.175$\,Hz/G$^2$ uncertainty of the quadratic Zeeman coefficient.
}
\label{fig:parabs}}
\end{figure}

Similar measurements were done for different values of the ``main'' bias magnetic field  $B^\textrm{m}_0$ in the range of $130-520$\,mG at the fixed value $B^\textrm{r}_0=218$\,mG.
Corresponding  data for  $\Delta\nu_{43}$, $\Delta\nu_{23}$, as well as for the synthetic frequency shift $\Delta\nu_\textrm{s}$, are shown in Fig.\,\ref{fig:parabs}a.
The frequency dependencies $\Delta\nu_{43}(B^\textrm{m}_0)$ and  $\Delta\nu_{32}(B^\textrm{m}_0)$ are
fitted by the parabolic function in the form $\beta  (B^m_{0}-B^r_{0})$, where $\beta$ is the quadratic Zeeman coefficient. 
We get $\beta_{43}=+257.47\,(22)(75)$\,Hz/G$^2$ and $\beta_{32}=-257.42\,(18)(75)$\,Hz/G$^2$. 
The value in the first parentheses represents $1\sigma$ statistical error of the fit. The value in the second parentheses corresponds to $1\sigma$ uncertainty due to the measurement error of the bias magnetic field (see Supplementary for the details).
The absolute values of $\beta_{43}$ and $\beta_{32}$ are equal within the combined uncertainty. 
The  averaged value $\beta = 257.44\,(14)(75)$\,Hz/G$^2$  agrees with the calculated value\cite{Golovizin2019inner} of  $\beta^\textrm{th} = 257.2$\,Hz/G$^2$.

The synthetic frequency shift depicted in  Fig.\,\ref{fig:parabs}c is also fitted by the parabolic function which gives  $\beta_\textrm{s}=-0.008(185)$\,Hz/G$^2$. It is compatible with zero within the measurement uncertainty and corresponds to at least 1000-fold reduction in sensitivity to the Zeeman shift compared to an individual clock transition. 

\textbf{Tensor Stark shift from the optical lattice field}

The Zeeman shift cancellation opens a way to  improve control on the tensor Stark shift. 
The tensor part of the clock transition frequency shift from the optical lattice is given by
\begin{equation}
    \label{eq:tensorshift}
    \Delta\nu^{t} =  \tilde{\alpha}^t\times (3 \cos^2\theta - 1) \times (U/E_\textrm{r}),
\end{equation}
where $\theta$ is the angle between the bias magnetic field $\Vec{B}_0$ and the lattice polarization vector $\Vec{\epsilon}$, as shown in Fig.\,\ref{fig:scheme}c. Here  $U$ is the lattice depth, $E_\textrm{r}=\hbar^2 k^2/(2m)=h\times 1043$\,Hz is the lattice photon recoil energy.
The coefficient $\tilde{\alpha}^t$ is associated with the differential tensor polarizability of $-0.20(4)$\,a.u. of the \clockw transition in Tm at 1064\,nm \cite{Golovizin2019inner} and is evaluated to be $0.7$\,Hz (see Supplementary). 
In our experiment, a 1D optical lattice at 1064\,nm is aligned vertically along the  \textit{z} axis, its polarisation $\Vec{\epsilon}$ is parallel to the  \textit{y} axis and $\Vec{B}_0$ is parallel to the \textit{x} axis, such that the  angle $\theta$ is close to $\pi/2$ (see Fig.\,\ref{fig:scheme}c).
Since $\Delta\nu^{t}$ has maximum for $\theta_0 = \pi/2$, its sensitivity to small fluctuation $\delta\theta$ around $\theta_0$ can be approximated by
\begin{equation}
\label{eq:tensorshiftvar}
  \delta^t=\Delta\nu^{t}(\pi/2+\delta\theta) - \Delta\nu^{t}(\pi/2)= 3\tilde{\alpha}^t  (U/E_\textrm{r})\times \delta\theta^2\,
\end{equation}
assuming  the constant value of $U$.
For the typical trap depth of  $U=100\,E_\textrm{r}$, accurate control of $\delta^{t}$ below 1\,mHz requires  $|\delta\theta| \lesssim 10^{-3}$. 
In our configuration $\delta\theta =  B_y/B_0$  assuming that the direction of lattice polarisation is much more stable than the direction of the external magnetic field. 
To reduce $\delta^{t}$ it is necessary to  increase the bias magnetic field $B_0$.  

\begin{figure}[]
\center{
\resizebox{0.5\textwidth}{!}{
\includegraphics{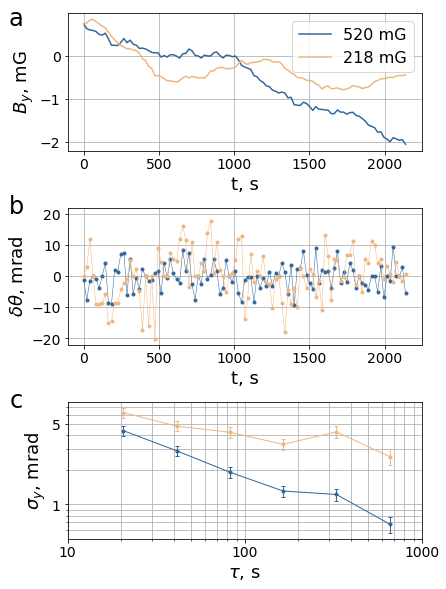}
}
\caption{Stabilization of the bias magnetic field direction. 
a) Compensation field  $B_y$ as a function of time for one measurement run. The orange line corresponds to the data points with  the ``reference'' field $B_0^\textrm{r}=218$\,mG, the blue line ---  to the points with ``main'' field $B_0^\textrm{m}=520$\,mG.
b) Evaluated $\delta\theta$ over the measurement run.
c) The Allan variance for $\delta\theta$ data sets from b). 
}
\label{fig:By_stab}}
\end{figure}

Using the bi-colour scheme we actively stabilize $\theta$ to  $\pi/2$ using the following method.
During a typical measurement run, we perform the perpendicular magnetic field calibration  every 5th cycle.
Clock transition frequency shifts $\delta_{\pm}^{t}$ are measured for two opposite signs of the deliberately added/subtracted  increment $\Delta\theta$ which gives $\theta = \pi/2 + \delta\theta \pm \Delta\theta$. The increment $\Delta\theta\approx 0.1$  is added by changing the perpendicular magnetic field $B_y$. This method significantly increases sensitivity of the frequency shift from $\sim\delta\theta^2$ dependency to $\sim\delta\theta\Delta\theta$. The misalignment $\delta\theta$ is deduced from the frequency mismatch  $\delta_+^{t} - \delta_-^{t}$ (see Supplementary for details).
Note, that  fluctuations of the magnetic field along $z$ axis do not influence angle~$\theta$.

Figure \ref{fig:By_stab}a shows traces of  perpendicular magnetic field $B_y$ produced by the compensation coils. For the given run we apply $B_0^\textrm{r}=218$\,mG and $B_0^\textrm{m}=520$\,mG. Depending on the value of the bias field (``reference'' or ``main''), different values of the compensation field are applied which are indicated as orange and blue lines, correspondingly. 
Plots  on Fig.\,\ref{fig:By_stab}b show the  misalignment angle $\delta\theta$ which is deduced from corresponding magnetic field measurements (see Supplementary).
Corresponding  Allan variance for $\delta\theta$ data sets is shown  in Fig.\,\ref{fig:By_stab}c. 
For larger bias magnetic field $B_0^\textrm{m}=520$\,mG, fluctuations of   $\delta\theta$  are smaller.  In this case the  uncertainty of  $\delta\theta = 1$\,mrad is achieved after 500\, seconds integration. For $B_0^\textrm{r}=218$\,mG  the averaging goes slower due to larger shot-to-shot fluctuations. 
Using these results one can evaluate contribution of the tensor Stark shift from the optical lattice (the lattice depth $U=300\,E_\textrm{r}$) using Eq.\,\ref{eq:tensorshiftvar}.
We get corresponding contribution to the fractional frequency uncertainty of  $\delta^{t}/\nu_0<10^{-17}$ for $B_0^\textrm{r}=218$\,mG and $\delta^{t}/\nu_0<10^{-18}$  for $B_0^\textrm{m}=520$\,mG after 500 seconds of integration. Note, that the increase of the bias magnetic field does not lead to additional Zeeman shift-associated uncertainty because of the bi-colour operation scheme.

\textbf{Discussion}

In this work we report on extraordinary low overall systematic frequency shifts in Tm optical clock operating at 1.14\,$\mu$m inner-shell magnetic dipole transition. In the bi-colour operation scheme we online measure the synthetic frequency $\nu_\textrm{s} = (\nu_{43} + \nu_{32})/2$ based on the simultaneous interrogation of two ground-state hyperfine components. 
We have proven  that the synthetic frequency allows to   cancel  out contribution of the Zeeman shift without any assumptions about magnetic field fluctuations. 
The experiments show that the Zeeman  shift of the synthetic frequency is consistent with zero within currently achievable statistical uncertainty, confirming the suppression of the quadratic Zeeman shift at least for 1000 times.
Using this technique we have experimentally demonstrated that  contribution of the  tensor Stark shift induced by the optical lattice can be also reduced to $10^{-18}$ level by accurate online tuning of the magnetic field direction.  Together with  results from  ref.\cite{Golovizin2019inner} demonstrating very low sensitivity of Tm clock transition to the BBR shift ($2.3\times10^{-18}$ at the room temperature), the suggested bi-colour Tm optical clock becomes almost free from  systematic frequency shifts at $10^{-18}$ level.  

The operational magic wavelength around  1064\,nm promises further advantages compared to many other optical lattice clocks. 
First, powerful and low-noise fiber lasers are available in this spectral region. 
Second, the sensitivity to the lattice wavelength is very low (1000 times smaller compared to Sr optical  clock), which results in  smaller lattice shifts from higher-order terms.
Summing up, bi-colour Tm optical clock operating at the 1.14\,$\mu$m inner-shell magnetic dipole transition demonstrates its functional capacity for the next generation of transportable systems with very moderate requirements to environmental conditions, temperature and magnetic field fluctuations. 

\section{Acknowledgments}

Authors acknowledge support from RSF grant \#19-12-00137. Authors are grateful to Denis Sukachev for careful reading and valuable discussion of the manuscript.
\bibliographystyle{unsrt}
\bibliography{references_mod, refs_new}

\appendix
\section*{The supplementary}
\textbf{The Zeeman shift and the synthetic frequency.}
The energy of an atomic level with the electronic momentum   $J$ and the nuclear spin   $I=1/2$ in an external magnetic field $B_0$ is given by \cite{Giglberger1967,Sukachev2016}:

\begin{equation}
\label{eq:BRmain}
\begin{split}
E_{J,F=J\pm1/2,m_F} = &-\frac{1}{4}hA_J + g_J \mu_B B_0 m_F  \\ 
&\pm \frac{h A_J (2J+1)}{4}\sqrt{1 - \frac{4 m_F}{2J+1}x + x^2},
\end{split}
\end{equation}
where $A_J$ is the hyperfine splitting constant, $x = \frac{2(g_J \mu_B - g_I \mu_N) B_0}{hA_J(2J+1)}$, $g_J$ and $g_I$ are the electronic and nuclear Land\'{e} g-factors, $\mu_B$ and $\mu_N$ are the Bohr and nuclear magnetons, correspondingly. 

For the magnetic sublevel with zero projection $m_F=0$ and $x \ll 1$ the level energy shift can be expressed as

\begin{equation}
\label{eq:BRsimple}
    \Delta E_{J,F=J\pm1/2} = \pm \frac{(g_J \mu_B - g_I \mu_N)^2 }{2hA_J(2J+1)}B_0^2 = \mp \beta_J B_0^2,
\end{equation}
where $\beta_J$ is the quadratic Zeeman shift coefficient for a given electronic level. 
The negative sign of $A_J$ for the ground and the clock levels in thuluim is taken into account.
The coefficients $\beta_{7/2}=426$\,Hz/G${}^2$ and $\beta_{5/2}=169$\,Hz/G${}^2$ are relatively large due to small (1.5\,GHz and 2.1\,GHz, respectively) hyperfine splitting of the ground and clock levels.
The frequencies of the \clocktransitionfirst and \clocktransitionsecond clock transitions as a function of $B_0$ can be found as

\begin{equation}
    \begin{split}
        h\Delta\nu_{43} &= \Delta E_{5/2,3} - \Delta E_{7/2,4} = -(\beta_{5/2} - \beta_{7/2}) B_0^2\,, \\
        h\Delta\nu_{32} &= \Delta E_{5/2,2} - \Delta E_{7/2,3} = (\beta_{5/2} - \beta_{7/2}) B_0^2.
    \end{split}
\end{equation}
Thus, for the synthetic frequency 
\begin{equation}
\label{eq:synthfreqshift}
    \Delta\nu_s = \frac{\Delta\nu_{43} + \Delta\nu_{32}}{2}
\end{equation}
the Zeeman shift completely cancels out. This result remains valid even without Taylor expansion of  Eq.\,\ref{eq:BRmain}. 

\textbf{The tensor polarizability coefficient.}
Following Ref.\cite{Golovizin2019inner}, an absolute value of the optical lattice trap depth can be expressed as
\begin{equation}
\label{eq:pol1}
    U = \left|-\alpha \frac{\mathcal{E}^2}{4}\right|\,,
\end{equation}
where $\mathcal{E}$ is the amplitude of the electric field and $\alpha=152$\,a.u. (atomic units) is the ground level polarizability.
The clock transition frequency shift from the tensor part of the polarizability is 
\begin{equation}
\label{eq:poldiff}
    h\Delta\nu^t = - \frac{3 \cos^2\theta - 1}{2} \Delta\alpha^t \frac{\mathcal{E}^2}{4}.
\end{equation}
Here $\Delta\alpha^t=-0.2$\,a.u.\cite{Golovizin2019inner} for the lattice wavelength of 1064\,nm.
Inserting Eq.\,\ref{eq:pol1} to Eq.\,\ref{eq:poldiff} and rearranging it, one gets
\begin{equation}
    \Delta\nu^t = -(3 \cos^2\theta - 1) \frac{U}{E_\textrm{r}} \times \frac{E_\textrm{r}}{2\alpha h}\Delta\alpha^t.
\end{equation}
Comparing this equation to Eq.\,\ref{eq:tensorshift} and using $E_\textrm{r} = h\times 1043$\,Hz we find
\begin{equation}
    \tilde{\alpha}^t = -\frac{E_\textrm{r}}{2 h}\frac{\Delta\alpha^t}{\alpha} = 0.7\,\textrm{Hz}
\end{equation}

\begin{figure}[t]
\center{
\resizebox{0.41\textwidth}{!}{
\includegraphics{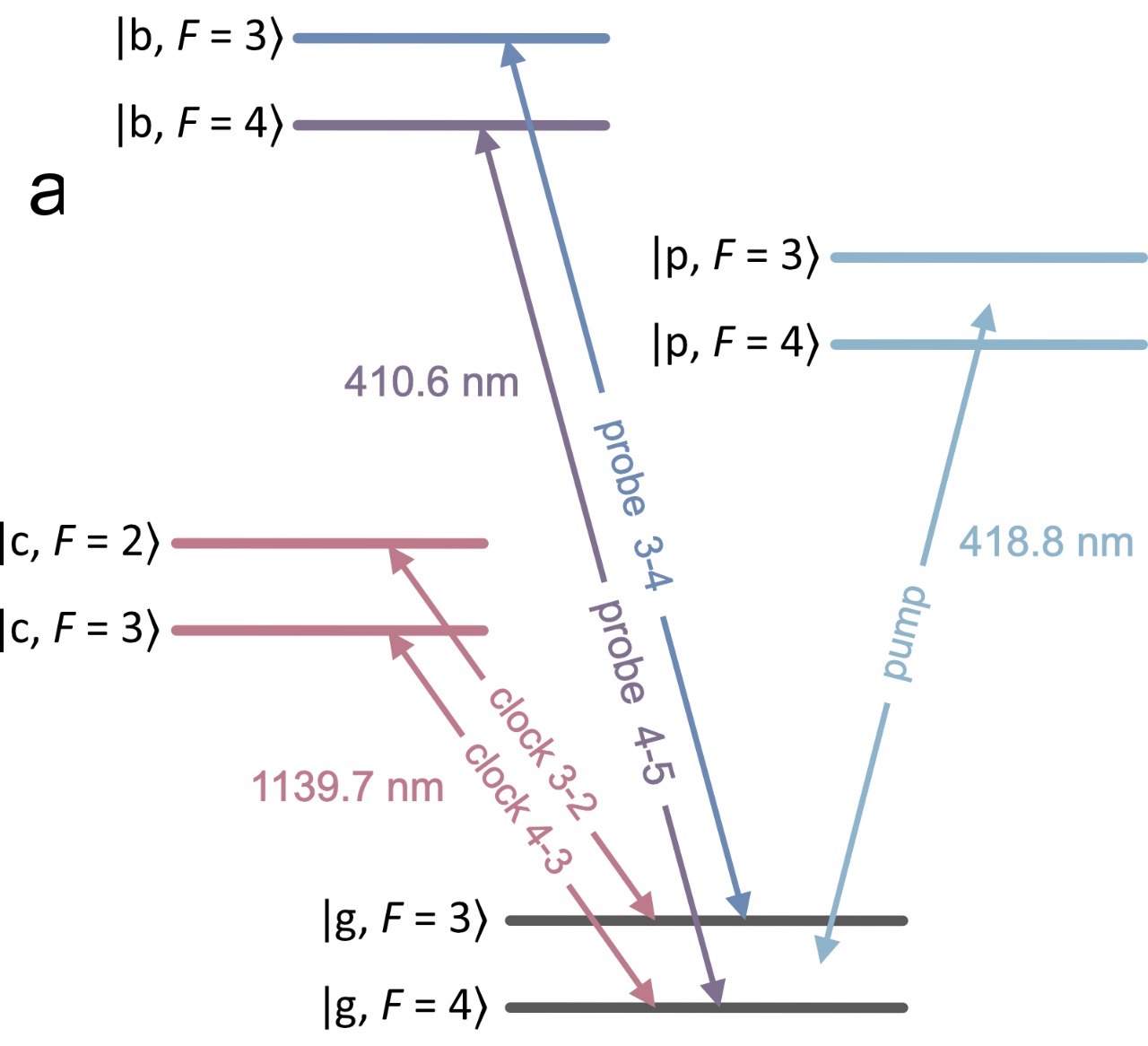}
}
\resizebox{0.57\textwidth}{!}{
\includegraphics{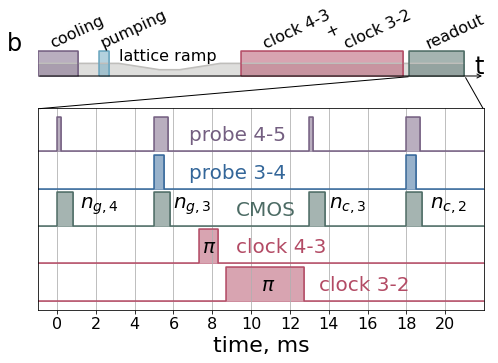}
}
\caption{
a) The relevant thulium level scheme. The ``pump'' radiation is detuned by $-175$\,MHz from $\ket{g,F=4}\rightarrow\ket{p,F=4}$ transition and by $-90$\,MHz from $\ket{g,F=3}\rightarrow\ket{p,F=3}$ transition.
b) The measurement cycle (top) and the detailed pulse sequence of the  readout procedure  (bottom). 
The ``CMOS'' line shows the exposition periods of the CMOS camera accumulating  the fluorescence signal  at 410\,nm. 
Labels $n_{\cdot,\cdot}$ indicate the signal recorded in the particular camera image. 
Clock 4-3 and clock 3-2 pulses are the $\pi-$pulses which transfer population from the upper clock levels to the ground levels.
}
\label{fig:readout}}
\end{figure}

\textbf{The excitation efficiency and the readout procedure.}
After simultaneous excitation of two clock transitions we implement a dedicated readout procedure  to deduce the  excitation efficiency of each of the transition.
Every readout is destructive, i.e. all measured atoms are removed from the trap. 
Below we use the  notation 
$\ket{b}$ for the $\ket{4f^{12}(^3H_5)5d_{3/2}6s^2,J=9/2}$ level, which is used for the first-stage laser cooling.
The relevant thulium levels are shown on Fig.\,\ref{fig:readout}a.
The sequence of the readout pulses is shown on Fig.\,\ref{fig:readout}b.

The readout procedure aims for the measurement of populations of four hyperfine sublevels $\ket{g,F=4}$, $\ket{g,F=3}$, $\ket{c,F=3}$ and $\ket{c,F=2}$ 
shown in  Fig\,\ref{fig:readout}a. The pulse sequence is the following:

\begin{enumerate}
    \item  with the help of $0.2$\,ms resonant probe\,4-5 pulse (\bluefirst) we measure the number of atoms remained in the $\ket{g,F=4}$ state after excitation by the  \clockfirstshort clock pulse, which is denoted as $n_{g,4}$;
    \item by two overlapped $0.7$\,ms resonant probe\,4-5  and $0.5$\,ms probe\,3-4 (\bluesecond) pulses we determine the number of atoms $n_{g,3}$ remained in the $\ket{g,F=3}$ state. The \bluesecond pulse repumps atoms from $\ket{g,F=3}$ to $\ket{g,F=4}$ state;
    \item two consecutive $\pi-$pulses  clock\,4-3 (1\,ms duration) and  clock\,3-2 ($4$\,ms duration)  return atoms from $\ket{c,F=3}$ to $\ket{g,F=4}$ and from $\ket{c,F=2}$ to $\ket{g,F=3}$ states, respectively; 
    \item similarly to p.1 we measure  $n_{c,3}$ which is proportional to the number of atoms excited by   clock\,4-3 pulse;
    \item similarly to p.2 we measure  $n_{c,2}$ which is proportional to the number of atoms excited by   clock\,3-2 pulse.
\end{enumerate}

\begin{figure}[t]
\center{
\resizebox{0.7\textwidth}{!}{
\includegraphics{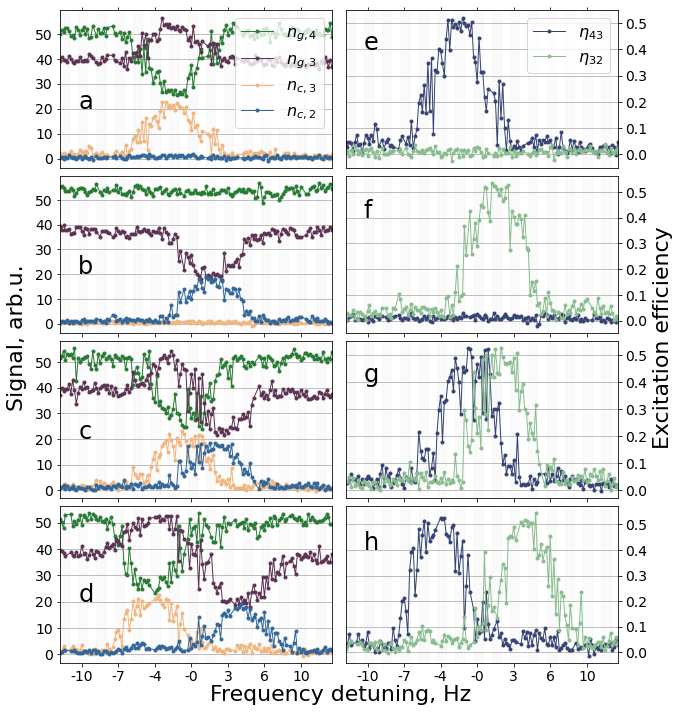}
}

\caption{
Simultaneous measurement of excitation  probabilities. 
The left column shows the raw data,  the right one shows excitation probabilities deduced from the corresponding raw data using  Eqs.\,\ref{eq:exprob43},\ref{eq:exprob32}.
The first row represents the case when 3-2 excitation field is switched off and only 4-3 transition is excited. The second row shows the case when 3-2 transition is excited solely.
In the third and the forth rows both transitions are  scanned using simultaneous excitation procedure. A deliberately introduced additional offset between two interrogating laser fields (3\,Hz in the 3rd row and 9\,Hz in the 4th row) define the time sequence of passing through two resonance curves. One can see that the resonance profiles are readily recovered both in the case of overlapped resonances and in the case of separate ones.
}
\label{fig:transcalib}}
\end{figure}

To accurately determine the excitation efficiencies, we deduce population of the corresponding clock levels  as following:
\begin{equation}
\begin{aligned}
    \tilde{n}_{g,4} &= n_{g,4}, \\
    \tilde{n}_{c,3} &= \xi_{c3} \times n_{c,3},\\
    \tilde{n}_{g,3} &= n_{g,3} - \xi_{g3} \times  n_{c,3},\\
    \tilde{n}_{c,2} &= \xi_{c2} \times n_{c,2}.
\end{aligned}    
\label{eq:calibration}
\end{equation}
Here the coefficients $\xi_{c3}$ and $\xi_{c2}$ are introduced to account for non-ideal $\pi$-pulses and spontaneous decay during the readout time from the clock $|c\rangle$ to the ground $|g\rangle$ states.
The coefficient $\xi_{g3}$ takes into account the  influence of spontaneous decay  from $\ket{c,F=3}$ to $\ket{g,F=3}$ and $\ket{g,F=4}$ during the time interval between the first and the second readout ``CMOS'' pulses (see Fig.\,\ref{fig:readout}b).
Note, that population of the  $\ket{c,F=2}$ level does not affect population of the $\ket{g,F=4}$ level, because the transition between them is forbidden.
Coefficients  $\xi_{g3}$, $\xi_{c3}$ and $\xi_{c2}$ are determined from the condition, that the total number of atoms associated with each transition $\tilde{n}_{g,4} + \tilde{n}_{c,3}$ and $\tilde{n}_{g,3} + \tilde{n}_{c,2}$ must not depend on the detuning of any of the clock pulses. 
Using this condition, the coefficients can be determined  from the individual scans of the ''4-3'' and the ''3-2'' clock transitions shown in Fig.\,\ref{fig:transcalib}a,b. 

Finally, the excitation probabilities of each of the clock transitions $\eta_{43}$ and $\eta_{32}$  can be calculated as:
\begin{align}
    \label{eq:exprob43}
    \eta_{43} &= \frac{\tilde{n}_{c,3}}{\tilde{n}_{g,4}+\tilde{n}_{c,3}} =  \frac{\xi_{c3}\, n_{c,3}}{n_{g,4}+\xi_{c3}\, n_{c,3}}\,,\\
    \eta_{32} &=\frac{\tilde{n}_{c,2}}{\tilde{n}_{g,3}+\tilde{n}_{c,2}} =  \frac{\xi_{c2}\,n_{c,2}}{n_{g,3} - \xi_{g3}\,n_{c,3} + \xi_{c2}\,n_{c,2}}\,.
    \label{eq:exprob32}
\end{align}

\textbf{Mutual influence of two clock transitions: excitation and readout.} 
Mutual influence of the clock transitions  may occur either  (i)  from the clock laser fields during the interrogation, or (ii) during the readout procedure. 

The first effect (i) is associated with the ac-Stark shift induced by the other interrogation field and equals   $\delta\nu^\textrm{AC} \approx \Omega^2/(4\pi^2\Delta\nu_\textrm{sep})=0.06\,\mu$Hz for a 80\,ms $\pi-$pulse with the Rabi frequency $\Omega$ and frequency difference $\Delta\nu_\textrm{sep}=617$\,MHz.   Other line pulling effects including  quantum interference are also negligible.

The second effect (ii) results from the population transfer between two pairs of clock levels and  is much more pronounced for  the 3-2 transition. First,  $\ket{g,F=3,m_F=0}$ level is initially 10 times less populated than $\ket{g,F=4,m_F=0}$. 
Second, the  upper  level $\ket{c,F=3,m_F=0}$ of the 4-3  transition decays to $\ket{g,F=3}$ with $1/28$ probability, while the transition from $\ket{c,F=2,m_F=0}$  to $\ket{g,F=4}$ is forbidden (Fig.\,\ref{fig:readout}a). 
The effect of population transfer is clearly seen in the raw data  in  Fig.\,\ref{fig:transcalib}a-d. 
However, if the timing of excitation and readout procedures remains unchanged, this effect is proportional to the number of atoms excited to the $\ket{c,F=3,m_F=0}$ level and can be eliminated using  Eq.\,\ref{eq:calibration}.

To qualitatively verify this assumption   we recorded  frequency scans of two resonances with strong overlapping (c,g) and small overlapping (d,h) as shown in Fig.\,\ref{fig:transcalib}. Generally, the scans of both transitions are independent and are defined by two frequency detunings from the certain $1.14\,\mu$m cavity mode. In this experiment we add a small fixed frequency offset between two interrogation laser fields  which defines corresponding scan windows and relative position of resonances in respect to the net frequency difference  of 617\,MHz. This small offset  impacts the relative population of  the upper clock levels $|c,\,F=2\rangle$ and $|c,\,F=3\rangle$ right  after  interrogation procedure. One can see that in both cases the spectral profiles of excitation probability are successfully recovered. 

Good agreement of the quadratic Zeeman frequency shift measured at different magnetic field values (Fig.\,\ref{fig:parabs}) also indicates that the readout procedure correctly recovers spectral profiles of the corresponding clock transitions. The quantitative estimation of the feasible impact of two digital locks is given below. 

\begin{figure}[b!]
\center{
\resizebox{0.4\textwidth}{!}{
\includegraphics{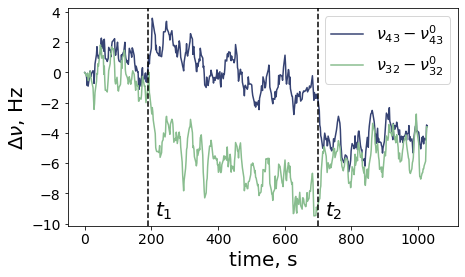}
}
\caption{Response of two parallel frequency lock channels to the instantaneous change of the bias field  $B_0$  from $218$\,mG to $231$\,mG at $t_1=190$\,s and back at $t_2=700$\,s. 
$\nu_{43}^0$ and $\nu_{32}^0$ are the corresponding AOMs frequencies at $t=0$. Laser frequency fluctuations and drift are common for both data sets.}
\label{fig:B0step}}
\end{figure}

\textbf{The digital lock performance.} Frequency locking of each of the interrogating light fields to the corresponding clock transition is performed by independent tuning of two  AOM frequencies as shown in  Fig.\,\ref{fig:scheme}c.
For every measurement cycle  we simultaneously excite both  clock transitions by the bi-colour radiation with each component alternatively detuned by $+\delta\nu/2$ or $-\delta\nu/2$ from the central frequency of the corresponding transition. Here  $\delta\nu$ is the measured transition linewidth which in our experiments is equal to $10$\,Hz. It  corresponds to the Fourier spectral width of a 80-ms interrogation pulse.
To measure the  Zeeman shift at two different magnetic fields $B_0^\textrm{r}$ and $B_0^\textrm{m}$  as described in the main text, we change magnetic field after probing excitation efficiencies on the left and right slopes (two consecutive cycles). Parallel  digital  lock  channels are responsible for two magnetic field values. 
The integral time constant of the digital proportional-integrating feedback loops equals $100$\,s.

As a test, we measured response of the digital locks to instantaneous change of the bias magnetic field $B_0=218$\,mG by a small value as shown in Fig.\,\ref{fig:B0step}.
At $t=190$\,s $B_0$ was increased by $\Delta B_0=13$\,mG, and at $t=700$\,s $B_0$ was returned to its original value.
Two transition frequencies become  shifted by the same amount in the opposite directions which corresponds to the estimated   Zeeman shift. 
The typical response time of the digital lock is of 10\,s.

\begin{figure}[b!]
\center{
\resizebox{0.36\textwidth}{!}{
\includegraphics{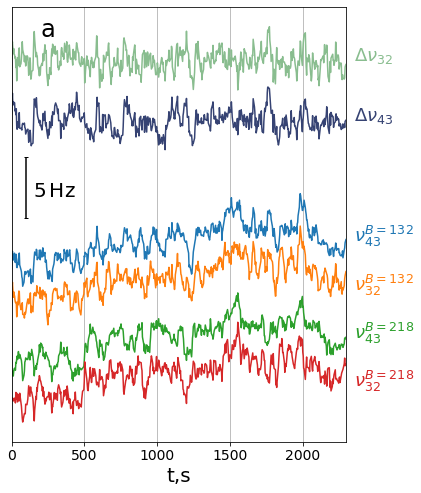}
}
\resizebox{0.305\textwidth}{!}{
\includegraphics{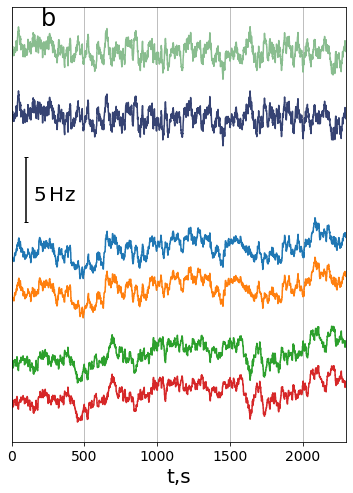}
}
\resizebox{0.21\textwidth}{!}{
\includegraphics{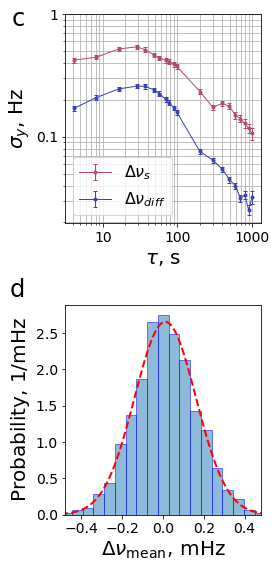}
}
\caption{Experimental frequency traces for $\Delta\nu_{43}$ and $\Delta\nu_{32}$ and individual traces $\nu_{43}(B^\textrm{m})$, $\nu_{32}(B^\textrm{m})$, $\nu_{43}(B^\textrm{r})$ and  $\nu_{32}(B^\textrm{r})$
for the bias magnetic fields $B^\textrm{r}_0=218$\,mG and  $B^\textrm{m}_0=132$\,mG (a) and results of simulations (b).
Each trace is  shifted for visual representation.
c) Comparison of the Allan variance plots (experiment) for the synthetic frequency $\Delta\nu_\textrm{s}=(\Delta\nu_{43}+\Delta\nu_{32})/2$ (red) and the differential frequency $\Delta\nu_\textrm{diff}=(\Delta\nu_{43}-\Delta\nu_{32})/2$ (blue).
d) The histogram of the difference between the 3-2 clock transition line centers from digital frequency locks with the 4-3 interrogation field ON and OFF (simulation). 3000 samples each corresponding for 1 hour data taking is used for the analysis.  
According to a Gaussian fit, the mean difference equals 12\,μHz with a standard deviation of 150\,μHz.
}
\label{fig:freqtraces}}
\end{figure}


To quantitatively evaluate the possible frequency shift from the readout procedure due to the population transfer between the levels we numerically simulated digital locks performance.
We generated a set of clock laser frequency signals with random frequency walk and white phase noise that is similar to the noise pattern of our $1.14\,\mu$m clock laser.
The white phase noise broadens the spectral linewidth to about  5\,Hz and long-term random frequency walk is usually within 100\,Hz during a day.
We repeated simulations for different frequency noise levels  including those that are considerably higher than in the experiment.
Simulation of the digital locks performance in the presence of the laser frequency fluctuations is shown in Fig.\,\ref{fig:freqtraces}b together with experimental data (Fig.\,\ref{fig:freqtraces}a).
Besides the intervals  $\Delta\nu_{43}$ and $\Delta\nu_{32}$  we show all four transition frequencies at two bias magnetic fields $B^\textrm{r}_0=218$\,mG and  $B^\textrm{m}_0=132$\,mG for each of the transition. The data comes from four corresponding  digital lock channels.  Experimental and simulated traces show  very similar behaviour.
Strong correlation between all four channels indicate that it is the laser frequency fluctuations which are mainly responsible for the short-time instability in Allan plots on figure \ref{fig:shifts}.
Comparison of two Allan variance plots for the experimental data on Fig.\,\ref{fig:freqtraces}c also demonstrates a $2.8$-times higher instability of the synthetic frequency $\Delta\nu_\textrm{s} = (\Delta\nu_{43} + \Delta\nu_{32})/2$ than of the differential frequency  $\Delta\nu_\textrm{diff} = (\Delta\nu_{43} - \Delta\nu_{32})/2$. Laser frequency noise should be strongly suppressed in  $\Delta\nu_\textrm{diff}$ which is proven by the data analysis. In turn, magnetic field fluctuations do not significantly contribute to the instability on the given level.

Using this noise model we also compare 3-2 clock transition line centres (deduced from the digital lock operation) for the case of 4-3 excitation radiation switched on and off. 
The histogram on Fig.\,\ref{fig:freqtraces}d consists of 3000 simulation runs each calculated for 1 hour-long data set which is similar to our experimental procedure. 
In total, the difference between two measurement schemes (4-3 field ON/OFF) equals  12\,μHz with a standard deviation of 150\,μHz. 
This proves that after reasonable  averaging time  population transfer during the readout does not impact uncertainty of the proposed clock scheme.

\textbf{Measurement of the bias magnetic field $B_0$.}
The magnitude of the bias magnetic field $B_0$ is deduced  from frequency measurements of two transitions $\ket{g,F=4,m_F=0}\rightarrow \ket{c,F=3,m_F=\pm1}$ possessing the strong first-order Zeeman sensitivity to the magnetic field. We use the formula
\begin{equation}
    \label{eq:sigmapm}
    B_0 = h\frac{\nu^+ - \nu^-}{2g_F\mu_B}\,,
\end{equation}
where $\nu^+$ and $\nu^-$ are the frequencies of $\sigma^{+}$ ($\ket{m_F=0}\rightarrow \ket{m_F'=1}$) and $\sigma^{-}$ ($\ket{m_F=0}\rightarrow \ket{m_F'=-1}$) transitions, respectively, $g_F=0.7121$ is Land\'e g-factor of the $\ket{c,F=3}$ level. 
In the experiment we successively record spectra of the  $\sigma^{+}$ and $\sigma^{-}$ transitions and determine  $\nu^+$ and $\nu^-$ frequencies (with respect to the frequency of the ULE cavity mode) from the approximations of the corresponding spectra with the Gaussian lineshape.
The typical frequency shift  $\nu^+ - \nu^-$ in our experiments was in the range from  $140$\,kHz to  $500$\,kHz. 
The observed FWHM of $\sigma^{\pm}$ transitions is about $2$\,kHz, which can be attributed to the fluctuations of the laboratory magnetic field. 
Typical statistical uncertainty of the line centre determination from the fit is $ 200$\,Hz. 
From Eq.\,\ref{eq:sigmapm} we estimate the net uncertainty  of the magnetic field determination to be $\sigma_{B_0}  = h \frac{\sqrt{2}\delta\nu}{2g_F\mu_B} = 0.14$\,mG.

\begin{figure}[]
\center{
\resizebox{0.4\textwidth}{!}{
\includegraphics{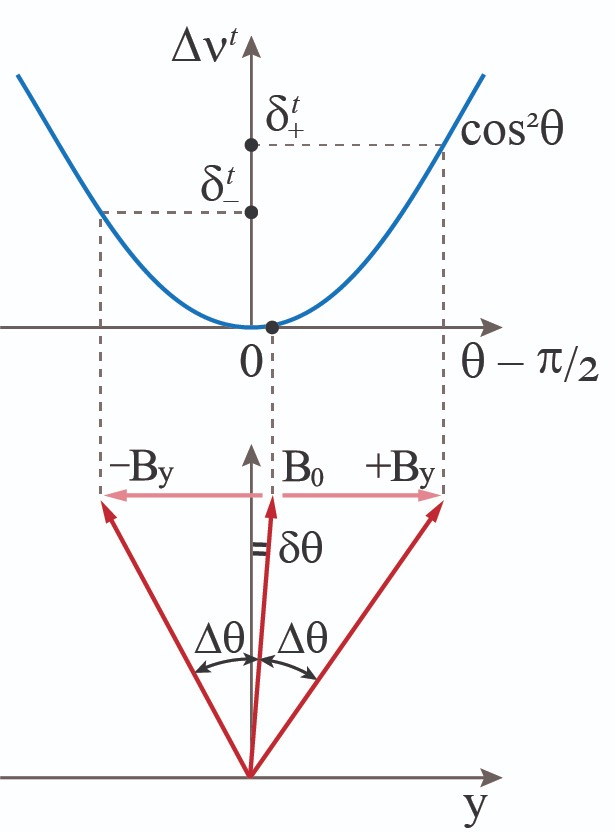}
}
\caption{Sketch of the bias magnetic field direction stabilization.
Top: clock 4-3 transition frequency shift as a function of misalignment angle from $\theta_0=\pi/2$.
Bottom: orientation of the bias magnetic field for nominal operation (small deviation $\delta\theta$  from $\theta_0=\pi/2$) and for calibration measurements $\pm\Delta\theta$. 
Rotation of $\vec{B}_0$ is accomplished by applying additional $\pm B_y$ during the clock interrogation period.
Misalignment angle $\delta\theta$ is deduced from the frequency difference $\delta^t_{\pm}$ according to Eq.\,\ref{eq:eq:tensorshiftaqq}.
}
\label{fig:B0alignment}}
\end{figure}

\textbf{Stabilization of the magnetic field direction.} 
To minimize sensitivity of the clock transitions frequencies (both 4-3 and 3-2) to the direction of the bias magnetic field $B_0$ we chose the angle $\theta=\pi/2$ (see Fig.\,\ref{fig:scheme} and Fig.\,\ref{fig:B0alignment}). 
According to Eq.\,\ref{eq:tensorshift}, the  sensitivity to angular deviation $\delta\theta$ at $\theta=\pi/2$ is quadratic.
Variations of the laboratory magnetic field can change the angle $\theta$, so one needs to implement an active stabilization. We use a pair of coils producing the field component $B_y$  which compensates corresponding component of the laboratory field.  

To minimize the deviation $\delta\theta$ we accomplish the following procedure individually for each of the  magnetic field values ($B^\textrm{m}_0$ and $B^\textrm{r}_0$). The procedure is  repeated after 5 regular measurement cycles.
We intentionally introduce additional angle  $\Delta\theta\approx0.1$ (the absolute value is not important here) between the field direction and the lattice polarization axis ($\vec\epsilon||y$) by changing $B_y$ at constant value of $B_x$ (Fig.\,\ref{fig:B0alignment}). To change $B_y$ we add/subtract   a fixed   current increment $\Delta I_y$ to the current flowing through the compensation coils.
Two transition frequencies $\nu_+^{t}$ and $\nu_-^{t}$ (for each of the fields   $B_0^{m}$, $B_0^{r}$) are measured for $\theta = \pi/2 + \delta\theta \pm \Delta\theta$ using a standard procedure of interrogating clock 4-3 transition on the left and the right slopes.
Frequency shifts $\delta_\pm^{t} = \nu_\pm^{t} - \nu_{43}$ are calculated with respect to the nominal transition frequency $\nu_{43}$ at $\theta=\pi/2 + \delta\theta$ measured in the last measurement cycle. 
Using Eq.\,\ref{eq:tensorshift}, $\delta\theta$  can be found as 
\begin{equation}
\label{eq:eq:tensorshiftaqq}
    \delta\theta = \frac{\delta_+^{t} - \delta_-^{t}}{12\tilde{\alpha}^t\Delta\theta\,(U/E_\textrm{r})}.
\end{equation}
For  $\Delta\theta = 0.1$ and $U=300\,E_\textrm{r}$
the target value of $|\delta\theta|=10^{-3}$ is reached for the  modest uncertainty of $|\delta_+^{t} - \delta_-^{t}| = 0.4$\,Hz, which with our current setup can be achieved in less than 500 cycles of measurement (see Fig.\,\ref{fig:shifts}b).
The single-measurement value $\delta\theta$ from Eq.\,\ref{eq:eq:tensorshiftaqq} is used as an error signal for the proportional-integrational digital lock, which tunes the offset value of $B_y$ (see Fig.\,\ref{fig:By_stab}a). 
Thus absolute values of $\Delta\theta$, as well as $\alpha^t$ and $U/E_r$, only affect the magnitude of the error signal.


\end{document}